\documentclass[12pt,onecolumn,draftclsnofoot,journal]{IEEEtran}

\usepackage[dvips]{graphicx}
\usepackage{cite}
\usepackage{amssymb}
\usepackage{amsmath}
\usepackage{amsthm}
\usepackage{multirow}
\usepackage{subfig}
\usepackage{color}
\usepackage{lineno}
\usepackage{enumerate}

\newtheorem{lemma}{Lemma}
\newtheorem{theorem}{Theorem}

\renewcommand{\IEEEQED}{\IEEEQEDopen}
\makeatletter

\newcommand{\Rmnum}[1]{\expandafter\@slowromancap\romannumeral #1@}

\newlength{\figa}
\makeatother

\begin{document}
\title{A Mathematical Proof of the Superiority of NOMA Compared to Conventional OMA} 
\author{Zhiyong~Chen,
        Zhiguo~Ding,~\IEEEmembership{Senior Member,~IEEE,}
        Xuchu~Dai,
        Rui~Zhang,~\IEEEmembership{Fellow,~IEEE}
        \thanks{Manuscript received XXX XX, 2016. The associate editor coordinating the review of this paper and approving it for publication was XXX.
        }
        \thanks{This work was supported in part by the National Natural Science Foundation of China (No. 61471334). The work of Z. Ding was supported by the Royal Society International Exchange Scheme and the UK EPSRC under grant number EP/N005597/1.}
        \thanks{ 
Z. Chen and X. Dai are with the Key Lab of Wireless-Optical Commun., Chinese Acad. of Sciences, Sch. of Info. Science \& Tech., Univ. Science \& Tech.  China, Hefei, Anhui, 230026, P. R. China. Z. Ding is with the School of Computing and Commun., Lancaster Univ., LA1 4YW, UK. (email: zhiyong@mail.ustc.edu.cn, z.ding@lancaster.ac.uk, daixc@ustc.edu.cn).        
        }
        \thanks{R. Zhang is with the Department of Electrical and Computer Engineering,
National University of Singapore, Singapore 117583, and also with the Institute
for Infocomm Research, A*STAR, Singapore (e-mail: elezhang@nus.edu.sg).}
		\thanks{Digital Object Identifier XXXX/XXXX
		}        
        }
\markboth{IEEE TRANSACTIONS ON SIGNAL PROCESSING,~Vol.~XX, No.~X, Oct.~2016}
{CHEN \MakeLowercase{\textit{et al.}}: A Mathematical Proof of the Superiority of NOMA Compared to Conventional OMA}

\maketitle
\begin{abstract}
While existing works about non-orthogonal multiple access (NOMA) have indicated that NOMA can yield a significant performance gain over orthogonal multiple access (OMA) with fixed resource allocation, it is not clear whether such a performance gain will diminish when optimal resource  (Time/Frequency/Power) allocation is carried out. 
In this paper, the performance comparison between  NOMA  and conventional OMA systems is investigated, from an optimization point of view. 
Firstly, by using the idea of power splitting, a closed-form expression for the optimum sum rate of NOMA systems is derived.
Then, with  rigorous mathematical proofs, we reveal the fact that  NOMA can always outperform  conventional OMA systems, even if both are equipped with   the optimal resource allocation policies.
Finally, computer simulations are conducted to validate the accuracy  of the analytical results.
\end{abstract}
\section{Introduction}
\IEEEPARstart{R}{ecently}, non-orthogonal multiple access (NOMA) has received extensive research interests  due to its superior spectral efficiency compared to conventional  orthogonal multiple access (OMA)~\cite{saito2013non,ding2014performance,benjebbour2013concept}. For example, NOMA has been proposed to downlink scenarios in 3rd generation partnership project long-term evolution (3GPP-LTE) systems \cite{3GPP}. Moreover,  NOMA has also been anticipated as a promising multiple access technique for the next generation cellular communication networks~\cite{li20145g,dai2015non}. 
 
Conventional multiple access techniques for cellular communications, such as frequency-division multiple access (FDMA) for the first generation (1G),
time-division multiple access (TDMA) for the second generation (2G), code-division multiple access (CDMA) used by both 2G and the third generation (3G), and orthogonal frequency division multiple access (OFDMA) for 4G, can all be categorized as OMA techniques, where different users are allocated to orthogonal resources, e.g., time, frequency, or code domain to avoid multiple access interference. However, these OMA techniques are far from the optimality, since that the spectrum resource allocated to the user with poor channel conditions cannot be efficiently used.

To tackle this issue and further improve spectrum efficiency, the concept of NOMA is proposed. 
The implementation of NOMA is based on the combination of superposition coding (SC) at the base station (BS) and successive interference cancellation (SIC) at users~\cite{saito2013non}, which can achieve the optimum performance for  degraded broadcast channels~\cite{cover2012elements,tse2005fundamentals}. Specifically,  take a two-user single-input single-output (SISO) NOMA
system as an example. The BS serves the users at the same time/code/frequency channel, where the signals are superposed 
with different power allocation coefficients. At the  user side, the
far user (i.e., the user with poor channel conditions) decodes its
message by treating the other's message as noise, while
the near user (i.e., the user with strong channel conditions) first
decodes the message of its partner and then decodes its own
message by removing partner's message from its observation. In this way, both users can have full access to all the resource blocks (RBs), moreover, the near user can decode its own
information without any interference from the far user. Therefore, the overall performance is enhanced, compared to conventional OMA techniques.  
\subsection{Related Literature}
As a promising multiple access technique, NOMA  and its variants have attracted considerable research interests recently.
The authors in \cite{saito2013non} firstly presented the concept of NOMA for cellular future radio access, and pointed out that NOMA can achieve higher spectral efficiency and better user fairness than conventional OMA. 
In \cite{ding2014performance}, the performance of NOMA in a cellular downlink scenario with randomly deployed users was investigated, which reveals that NOMA can achieve superior performance in terms of ergodic sum rates. In \cite{ding2015cooperative}, a cooperative NOMA scheme was proposed by fully exploiting prior information  at the users with strong channels about the messages of the users with weak channels. The impact of user pairing on the performance of NOMA systems was characterized in \cite{ding2014impact}. In \cite{xu2015new}, a new evaluation criterion was developed to investigate the performance of NOMA, which shows that NOMA can outperform OMA in terms of the sum rate, from an information-theoretic point of view. 

To further improve spectral efficiency, the combination of NOMA and multiple-input multiple-output (MIMO) techniques, namely MIMO-NOMA, has also been extensively investigated.
In \cite{Ding16Application}, a new design of precoding and detection matrices for MIMO-NOMA was proposed. A novel MIMO-NOMA framework for downlink and uplink transmission was proposed by applying the concept of signal alignment in \cite{Ding16General}. To characterize the performance gap between MISO-NOMA and optimal dirty paper coding (DPC), a novel concept termed quasi-degradation for multiple-input single-output (MISO) NOMA downlink was introduced in \cite{Chen2016Opt}. Then, the theoretical framework of quasi-degradation was fully established in \cite{Chen.TSP}, including the mathematical proof of the properties, necessary and sufficient condition, and occurrence probability.  Consequently, practical algorithms for multi-user downlink MISO-NOMA systems were proposed in \cite{Chen.Access}, by taking advantage of the concept of quasi-degradation.
Lately, to optimize the overall bit error ratio (BER) performance of MIMO-NOMA downlink, an interesting transmission scheme  based on minimum Euclidean distance (MED) was proposed in \cite{Chen.TVT2016}.
\subsection{Contributions}
While existing works about NOMA have indicated that NOMA can yield a significant performance gain over OMA with fixed resource allocation, it is not clear whether such a performance gain will diminish when optimal resource allocation is carried out. 
In this paper, the performance comparison between NOMA and OMA is evaluated, from an optimization point of view, where optimal resource allocation is carried out to both multiple access schemes. 
In this paper, two kinds of OMA systems are considered, i.e., OMA-TYPE-I and OMA-TYPE-II, which represent, respectively, OMA systems with optimum power allocation and fixed time/frequency allocation, and OMA systems with both optimum power and time/frequency allocation.
The contributions of this paper can be summarized as follows.
\begin{enumerate}
\item The optimization problems for both NOMA and OMA systems are formulated, with consideration of user fairness. Particularly, more sophisticated OMA systems with joint power and time/frequency optimized are also considered. 
\item The closed-form expression  of the optimum  sum rate for NOMA systems is given, by taking advantage of the power splitting method.
\item By introducing the minimum required power for different systems, it is pointed out that the minimum required power of NOMA is always smaller than that of both OMA-TYPE-I and OMA-TYPE-II systems.
\item It is revealed that the optimum sum rate of NOMA systems is always larger than  that of both OMA-TYPE-I and OMA-TYPE-II systems, with various user fairness considerations, by  rigorous  mathematical proofs. 
\end{enumerate}

\subsection{Organization}
The remainder of this paper is organized as follows. Section II briefly describes the system model and the problem formulation. Section III provides the optimal power allocation policies as well as their performance comparison. Simulation results are given in Section IV, and Section V summarizes this paper.

\section{Problem Formulation}
Consider a downlink communication system with one BS and $K$ users, where   the BS and all the users are equipped with a single antenna. By using NOMA transmission, the received signal at user $i$ is
\begin{equation}
\label{eq:NOMA_model}
y=h_ix+n_i,\quad i=1,2,...,K,
\end{equation}
where $h_i$ denotes the channel coefficient, and $n_i\sim \mathcal{CN}(0,N_0)$ is the additive white Gaussian noise (AWGN) at user $i$. 
$x=\sum_{i=1}^K\sqrt{P_i}s_i$ is the superposition   of $s_i$'s with power allocation policy $\mathcal{P}=\{(P_1,P_2,...,P_K)|\sum_{i=1}^KP_i=P\}$,   $s_i$ represents the data intended to convey to user $i$, $P_i$ denotes the power allocated to user $i$, and $P$ denotes the total power constraint. For ease of analysis, we assume that $|h_1|\geq |h_2|\geq ... \geq |h_K|$ and the total bandwidth is normalized to unity in this paper.

In consideration of user fairness, herein, we introduce the minimum rate constraint $r^*$. Mathematically, the power allocation policy should guarantee the following constraint:
\[
\min_i r_i\geq r^*,
\] 
where $r_i$ is the achievable rate of user $i$ in nats/second/Hz, which is given by
\begin{equation}\label{eq:ri_NOMA}
r_i=\ln \Big(1+\frac{P_i|h_i|^2}{N_0+|h_i|^2\sum_{j=1}^{i-1}P_j}\Big).
\end{equation}
For the special case of  $i=1$, the summation in the denominator becomes $0$, and the corresponding rate becomes 
\[
r_1=\ln \Big(1+\frac{P_i|h_i|^2}{N_0}\Big).
\]
Note that $r_i$ is achievable since the channels are ordered and the user with strong channels can decode those messages sent to the users with weaker channels.

Therefore, the optimization problem of maximizing the total sum rate with the user fairness constraint for NOMA systems can be formulated as follows:
\begin{equation}\label{eq:Opt_NOMA} 
\begin{aligned}
R_{ {N}}\triangleq \max_{{P}_i}\quad &\sum_{i=1}^K r_i \\
\mathrm{s.t.} \quad     &r_i=\ln \Big(1+\frac{P_i|h_i|^2}{N_0+|h_i|^2\sum_{j=1}^{i-1}P_j}\Big),\\
						&\sum_{i=1}^K P_i\leq P,\\
                        &\min_i r_i\geq r^*.
\end{aligned}
\end{equation}

In traditional OMA systems, e.g., frequency division multiple access (FDMA) or time division multiple access (TDMA),   time/frequency resource allocation is non-adaptively fixed, i.e., each user is allocated with a fixed sub-channel. For notational simplicity, we refer to this type of OMA as OMA-TYPE-I in this paper.
Consequently, to optimize the power allocations, the optimization problem of OMA-TYPE-I assuming equal resource (time or frequency) allocation to all users  can be formulated as follows:
\begin{equation}\label{eq:Opt_OMA} 
\begin{aligned}
R_{ {O1}}\triangleq\max_{{P}_i}\quad &\sum_{i=1}^K r_i\\
\mathrm{s.t.} \quad     &r_i=\frac{1}{K}\ln \Big(1+\frac{KP_i|h_i|^2}{N_0} \Big),\\
						&\sum_{i=1}^K P_i\leq P,\\
                        &\min_i r_i\geq r^*.\\                    
\end{aligned}
\end{equation}

Since the sub-channel allocations among users are not optimized, some users may suffer from poor channel conditions due to large path loss and random fading. Thus, the optimization problem for jointly designing power and sub-channel allocations is considered next. Specifically, the total time/frequency is divided into $N$ sub-channels to be orthogonally shared by $K$ users, and this optimization problem can be formulated as follows:
\begin{equation}\label{eq:Opt_OMA_JO_Ex} 
\begin{aligned}
R_{ {OX}}\triangleq \max_{{P}_{i,n},S_i}\quad &\sum_{i=1}^K\sum_{n\in S_i} r_{i,n}\\
\mathrm{s.t.} \quad     &r_{i,n}={\frac{1}{N}}\ln \Big(1+\frac{P_{i,n}|h_{i,n}|^2}{N_0\frac{1}{N}} \Big),\\
						&\sum_{n=1}^N\sum_{i=1}^K P_{i,n}\leq P,\\
						&P_{i,n}\geq 0,\quad \forall i,n\\
                        &\sum_{n\in S_i} r_{i,n}\geq r^*,\\
                        &S_1,S_2,...,S_K\quad \text{are disjoint},\\
                        &S_1\cup S_2\cup ... \cup S_K = \{1,2,...,N\},
\end{aligned}
\end{equation}
where $P_{i,n}$ and $h_{i,n}$ are the power allocated to and the channel coefficient of user $i$'s sub-channel $n$, respectively. $S_i$ is the set of indices of sub-channels assigned to user $i$. 

Note that the optimization problem in \eqref{eq:Opt_OMA_JO_Ex} is not a convex problem. Fortunately, it is observed that it can be upper-bounded by the following optimization problem by replacing the discrete time/frequency allocation with a continuous one as follows:
\begin{equation}\label{eq:Opt_OMA_JO} 
\begin{aligned}
R_{ {O2}}\triangleq\max_{{P}_i,\alpha_i}\quad &\sum_{i=1}^K r_i\\
\mathrm{s.t.} \quad     &r_i={\alpha_i}\ln \Big(1+\frac{P_i|h_i|^2}{\alpha_i N_0} \Big),\\
						&\sum_{i=1}^K P_i\leq P,\\
                        &\min_i r_i\geq r^*,\\
                        &\sum_{i=1}^{K} \alpha_i =1.
\end{aligned}
\end{equation}
For notational simplicity, in  this paper , we refer to the OMA system with the optimization given in \eqref{eq:Opt_OMA_JO} as OMA-TYPE-II.

Note that the optimization problems in \eqref{eq:Opt_OMA} and \eqref{eq:Opt_OMA_JO} are applicable to both TDMA  and FDMA, due to the fact that over all user orthogonal time slots  the energy conservation $\sum_{i=1}^K\alpha_i\frac{P_i}{\alpha_i}=P$ is established in TDMA and the effective noise power becomes ${\alpha}{N_0}$ in FDMA.

By observing the definitions of the three kinds of OMA systems, it is implied that
\[
R_{O1}\leq R_{OX}\leq R_{O2}.
\]
Therefore, to show the superiority of NOMA compared to OMA, we only need to prove that
\[
R_N\geq R_{O2}.
\]
However, to dig out more sophisticated properties of these OMA systems, OMA-TYPE-I and OMA-TYPE-II are both considered in this paper. Moreover, different mathematical skills need to be employed to prove the superiority of NOMA compared to OMA-TYPE-I and OMA-TYPE-II,  respectively.
 
\section{Optimal Performance Analysis}
\subsection{Closed-form Solution of NOMA}
The optimum closed-form solution of NOMA is given in Theorem \ref{tm:1}.
\begin{theorem}
\label{tm:1}
Given $P$ and $r^*$, if 
\begin{equation}
\label{eqTM:P_N^*}
P_N^* \triangleq (e^{r^*}-1)N_0\sum_{i=0}^{K-1}\frac{e^{ir^*}}{|h_{K-i}|^2}\leq P,
\end{equation}

then,  the optimization problem in \eqref{eq:Opt_NOMA} is feasible, and the optimal solution can be written as
\begin{equation}
\label{eq:NOMA_sol}
R_{ {N}}=Kr^*+\Delta r_N,
\end{equation}
where
\begin{equation}\label{eq:NOMA_sol_1}
\Delta r_N=\ln\Big(1+\frac{(P-P_N^* )|{h_1}|^2}{N_0e^{Kr^*}}\Big).
\end{equation}
\end{theorem}
\begin{IEEEproof}
Following the idea introduced in~\cite{jindal2003capacity},
we split the total power into two parts, 1) the minimum power for supporting the minimum rate transmission, denoted by $P_N^*$, 2) the excess power, denoted by $\Delta P_N $. Denote the minimum power for maintaining minimum rate transmission and the excess power of user $i$ by $P^*_i$ and $\Delta P_i$, respectively. The minimum power $P^*_i$ is defined as follows.
If all users are allocated their minimum powers, then all users will achieve the minimum rate. Mathematically,  $P_i^*$ is defined as
\begin{equation}
\label{eq:tm1_r^*} 
r^*=\ln\Big( 1+\frac{P_i^*}{\frac{N_0}{|h_i|^2}+\displaystyle{\sum_{j<i}}P_j^*}\Big).
\end{equation}

Then, we have the following equalities.
\begin{equation}\label{eq:tm1_4eq} 
\left\{
\begin{aligned}
&P_i=P^*_i+\Delta P_i, \quad
P_N^*=\sum_{i=1}^{K} P^*_i,\\
&\Delta P_N=\sum_{i=1}^{K} \Delta P^*_i,\quad
P=P_N^*+\Delta P_N.
\end{aligned}
\right. 
\end{equation}

%
It follows from the definition that the minimum power of each user can be given by 
\begin{equation}
\label{eq:T1_P_i^*} 
P_i^*=(e^{r^*}-1)\Big(\frac{N_0}{|h_i|^2}+\displaystyle{\sum_{j<i}}P_j^*\Big).
\end{equation}
Therefore, we can obtain the following expression for the sum power of the minimum power $P_i^*$
\begin{equation}
\label{eq:P_N^*} 
P_N^*=\sum_{i=1}^K P_i^*=(e^{r^*}-1)N_0\sum_{i=0}^{K-1}\frac{e^{ir^*}}{|h_{K-i}|^2}.
\end{equation}
By combining \eqref{eq:tm1_r^*} and \eqref{eq:T1_P_i^*}, the minimum rate $r^*$ can also be written as
\begin{equation}\label{eq:T1_r^*} 
r^* =\ln\Big( 1+\frac{P_i^*+(e^{r^*}-1)\displaystyle{\sum_{j<i}}\Delta P_j}{\frac{N_0}{|h_i|^2}+\displaystyle{\sum_{j<i}}(P_j^*+\Delta P_j)}\Big).
\end{equation}
Then, the rate increment for user $i$ can be calculated as
\begin{equation}\label{eq:deltaR_2} 
\begin{aligned}
\Delta r_i
&=\ln\Big(1+\frac{P_i^*+\Delta P_i }{\frac{N_0}{|h_i|^2}+\displaystyle{\sum_{j<i}}(P_j^*+\Delta P_j)}\Big)-r^*\\
&=\ln\Bigg(1+\frac{\Delta P_i -(e^{r^*}-1)\displaystyle{\sum_{j<i}}\Delta P_j }{\frac{N_0}{|h_i|^2}+\displaystyle{\sum_{j\leq i}}P_j^* + e^{r^*} \displaystyle{\sum_{j<i}}\Delta P_j}\Bigg).
\end{aligned}
\end{equation}
By defining
\begin{equation*}\label{eq:de_Pie} 
\left\{
\begin{aligned}
&  P_i^e=\big(\Delta P_i -(e^{r^*}-1)\displaystyle{\sum_{j<i}}\Delta P_j \big)e^{(K-i)r^*},\\
&n_i^e=\big(\frac{N_0}{|h_i|^2}+\displaystyle{\sum_{j\leq i}}P_j^*\big)e^{(K-i)r^*},
\end{aligned}
\right. 
\end{equation*}
we have
\begin{equation}
\label{eq:delta_r3} 
\Delta r_i=\ln\Big(1+\frac{P_i^e}{n_i^e+\displaystyle{\sum_{j<i}}P_j^e}\Big).
\end{equation}

Consequently, the optimization problem in \eqref{eq:Opt_NOMA} can be equivalently written as
\begin{equation}\label{eq:Opt_NOMA_eq} 
\begin{aligned}
\max_{\mathcal{P}}\quad &Kr^*+\sum_{i=1}^K \Delta r_i \\
\mathrm{s.t.} \quad     &\sum_{i=1}^K P_i^e\leq P-P_N^*,\\
                        &\Delta r_i=\ln \Big(1+\frac{P_i^e}{n_i^e+  \displaystyle{\sum_{j<i}}P_j^e}\Big).
\end{aligned}
\end{equation}
The solution of \eqref{eq:Opt_NOMA_eq} is trivial. It is optimal to allocate all the  power to user $1$, i.e., the user with the strongest channel condition. Thus, the excess rate at user $1$ is 
\begin{equation}\label{eq:delta_r4} 
\begin{aligned}
\Delta r_1&=\ln \Big(1+\frac{P-P_N^*}{n_1^e}\Big)\\
          &=\ln \Big(1+\frac{P-P_N^*}{(\frac{N_0}{|h_1|^2}+P_1^*)e^{(K-1)r^*}}\Big)\\
          &=\ln \Big(1+\frac{(P-P_N^*)|h_1|^2}{  N_0 e^{Kr^*}}\Big),
\end{aligned}
\end{equation}
and the excess rates at other users are all $0$.
In other words, the excess sum rate is
\[
\Delta r_N =\Delta r_1=\ln \Big(1+\frac{(P-P_N^*)|h_1|^2}{  N_0 e^{Kr^*}}\Big),
\]
and the proof is complete.
\end{IEEEproof}

\subsection{Solution of OMA-TYPE-I}
The superiority of NOMA compared to OMA-TYPE-I is shown in Theorem \ref{tm:2}.
\begin{theorem}
\label{tm:2}
Given $P$ and $r^*$, 
if 
\begin{equation}
\label{eqTM:P_O1^*}
P_{O1}^* \triangleq (e^{Kr^*}-1)\frac{N_0}{K}\sum_{i=1}^{K }\frac{1}{|h_{i}|^2} \leq P,
\end{equation}

then, the optimization problem in \eqref{eq:Opt_OMA} is feasible,   the optimal solution must satisfy 
\[
R_{O1}\leq R_N,
\]
and the equality holds only when  $|h_1|=|h_2|=...=|h_K|$.
\end{theorem}

\begin{IEEEproof}
Similar as the proof of Theorem \ref{tm:1}, to obtain the solution of the optimization problem in \eqref{eq:Opt_OMA}, the total power is split into two parts, i.e., the minimum power for supporting minimum rate transmission, and the excess power. 

For user $i$, it is noted that the minimum power $P_i^*$ should satisfy
\[
\frac{1}{K}\ln\Big(1+\frac{KP_i^*|h_i|^2}{N_0}\Big)=r^*.
\] 
Hence, we can obtain
\[
P_i^*=(e^{Kr^*}-1)\frac{N_0}{K}\frac{1}{|h_i|^2},
\]
and the total minimum power $P_{O1}^*$ can consequently be written as 
\[
P_{O1}^*=\sum_{i=1}^KP_i^*=(e^{Kr^*}-1)\frac{N_0}{K}\sum_{i=1}^K\frac{1}{|h_i|^2}.
\]

On the other hand, given user $i$, the rate increment with excess power $\Delta P_i$ can be calculated as
\begin{equation*}\label{eq:*1}
\begin{aligned}
\Delta r_i&=\frac{1}{K}\ln\Big(1+\frac{K(P_i^*+\Delta P_i)|h_i|^2}{N_0}\Big)-r^*\\
          &=\frac{1}{K}\ln\Big(1+\frac{K\Delta P_i|h_i|^2}{N_0+KP_i^*|h_i|^2}\Big)\\
          &=\frac{1}{K}\ln\Big(1+\frac{K\Delta P_i}{N_0}\frac{|h_i|^2N_0}{N_0+KP_i^*|h_i|^2}\Big)\\
          &=\frac{1}{K}\ln\Big(1+\frac{K\Delta P_i}{N_0}|h_i|^2e^{-Kr^*}\Big).\\
\end{aligned}
\end{equation*}
By defining
\[
|\bar{h_i}|^2\triangleq  |h_i|^2 e^{-Kr^*},
\]
the rate increment can be simply written as 
\[
\Delta r_i=\frac{1}{K}\ln\Big(1+\frac{K\Delta P_i|\bar{h_i}|^2}{N_0}\Big).
\]

Therefore, the optimization problem in \eqref{eq:Opt_OMA} can be transformed to the problem as follows:
\begin{equation}\label{eq:Opt_OMA_eq} 
\begin{aligned}
R_{O1}=\max_{\mathcal{P}}\quad &Kr^*+\sum_{i=1}^K \Delta r_i\\
\mathrm{s.t.} \quad     &\sum_{i=1}^K P_i\leq P-P_{O1}^*,\\
                        &\Delta r_i=\frac{1}{K}\ln \Big(1+\frac{KP_i|\bar{h_i}|^2}{N_0} \Big) .
\end{aligned}
\end{equation}
It is well known that, the optimal solution can be obtained by the water-filling power allocation policy \cite{boyd2004convex}. Specifically, the optimal solution can be written as
\begin{equation}
\label{eq:OMA_sol}
R_{O1}=Kr^*+\Delta r_{O1},
\end{equation}
where
\begin{equation}\label{eq:OMA_sol_1} 
\left\{
\begin{aligned}
&\Delta r_{O1}=\frac{1}{K}\sum_{i=1}^K\ln\Big( \frac{K|\bar{h_i}|^2}{N_0}\mu\Big)\mathbf{ 1}\Big(\mu>\frac{N_0}{K|\bar{h_i}|^2}\Big),\\
&\sum_{i=1}^K\Big[\mu- \frac{N_0}{K|\bar{h_i}|^2}\Big]^+=P-P_{O1}^*.
\end{aligned}
\right. 
\end{equation}
Here, $[x]^+ \triangleq \max(x,0)$, and $\mathbf{1}()$ denotes the indicator function.

On the other hand, it is noted  that $\Delta r_{O1}$ can be alternatively represented as
\begin{equation}\label{eq:Deltar_O} 
\begin{aligned}
\Delta r_{O1}=\max_\mathcal{P}\quad &\sum_{i=1}^K\frac{1}{K}\ln \Big(1+\frac{KP_i|\bar{h_i}|^2}{N_0}\Big)\\
\mathrm{s.t.}\quad &\sum_{i=1}^KP_i\leq P-P_{O1}^* .
\end{aligned}
\end{equation}
By using the Arithmetic Mean-Geometric Mean (AM-GM) inequality, we have
\begin{equation}\label{eq:tm3_Deltar_O} 
\begin{aligned}
&\sum_{i=1}^K\frac{1}{K}\ln \Big(1+\frac{KP_i|\bar{h_i}|^2}{N_0}\Big)\\
&= \ln \prod_{i=1}^K \Big(1+\frac{KP_i|\bar{h_i}|^2}{N_0}\Big)^\frac{1}{K}\\
&\leq \ln \frac{1}{K} \sum_{i=1}^K \Big(1+\frac{KP_i|\bar{h_i}|^2}{N_0}\Big)\\
&=\ln \Big(1+\sum_{i=1}^K \frac{P_i|\bar{h_i}|^2}{N_0}\Big).
\end{aligned}
\end{equation}
The equality holds when
\begin{equation}
\label{eq:tm3_e2} 
 |\bar{h_1}|=|\bar{h_2}|=...=|\bar{h_K}|.
\end{equation}

By combining \eqref{eq:Deltar_O} and \eqref{eq:tm3_Deltar_O}, we can obtain
\begin{equation}\label{eq:tm3_rO_upper} 
\begin{aligned}
\Delta r_{O1}\leq \max_\mathcal{P}\quad & \ln \Big(1+\sum_{i=1}^K \frac{P_i|\bar{h_i}|^2}{N_0}\Big)\\
\mathrm{s.t.}\quad &\sum_{i=1}^KP_i\leq P-P_{O1}^* .
\end{aligned}
\end{equation}
The optimal solution of the optimization problem in \eqref{eq:tm3_rO_upper} is to allocate all the power to user 1, i.e., $P_1=P-P_{O1}^*$.
Therefore, we can have
\begin{equation}\label{eq:TM2_Part1} 
\Delta r_{O1}\leq \ln \Big(1+  \frac{(P-P_{O1}^*)|\bar{h_1}|^2}{N_0}\Big).
\end{equation}


Here, we introduce the following basic inequality.
\begin{lemma}[Chebyshev's Sum Inequality]\label{lm:1}
Let $a_1\geq a_2\geq...\geq a_K$ and $b_1\geq b_2\geq...\geq b_K$  be {strictly positive} numbers. Then
\[
\sum_{i=1}^Ka_ib_i\geq \frac{1}{K}\sum_{i=1}^K a_i\sum_{i=1}^{K} b_i\geq \sum_{i=1}^Ka_ib_{K+1-i}.
\]
The two inequalities become equalities when $a_1=a_2=...=a_K$ or $b_1=b_2=...=b_K$.
\end{lemma} 
By using Lemma \ref{lm:1}, we have
\begin{equation}\label{eq:TM2_Part2} 
\begin{aligned}
P_N^*&=(e^{r^*}-1)N_0\sum_{i=0}^{K-1}\frac{e^{ir^*}}{|h_{K-i}|^2}\\
     &\leq (e^{r^*}-1)N_0 \frac{1}{K}\sum_{i=0}^{K-1} e^{ir^*}\sum_{i=0}^{K-1}\frac{1}{|h_{K-i}|^2}\\
     &=(e^{Kr^*}-1)\frac{N_0}{K}\sum_{i=1}^{K }\frac{1}{|h_i|^2}\\
     &=P_{O1}^*.
\end{aligned}
\end{equation}
The equality holds when 
\begin{equation}
\label{eq:tm3_e1}
r^*=0 \quad \mathrm{or} \quad |h_1|=|h_2|=...=|h_K|.
\end{equation}
By the definition of $|\bar{h_i}|^2$, we have
\begin{equation}
\label{eq:TM2_Part3} 
 |\bar{h_1}|^2={|h_1|^2}e^{-Kr^*} .
\end{equation}
By combining the inequalities in \eqref{eq:TM2_Part1} and \eqref{eq:TM2_Part2} and equality in \eqref{eq:TM2_Part3}, we can have
\begin{equation}\label{eq:tm3_b} 
\begin{aligned}
\Delta r_{O1} &\leq \ln \Big(1+  \frac{(P-P_O^*)|\bar{h_1}|^2}{N_0}\Big)\\ 
 &\leq \ln \Big(1+  \frac{(P-P_N^*)|\bar{h_1}|^2}{N_0}\Big)\\
&= \ln \Big(1+  \frac{(P-P_N^*)|{h_1}|^2}{N_0e^{Kr^*}}\Big)\\
& = \Delta r_N.
\end{aligned}
\end{equation}
It is also worth noting that 
the first inequality  becomes equality when 
\[|h_1|=|h_2|=...=|h_K|,\] 
and the second inequality  becomes equality when
\[r^*=0\quad \mathrm{or} \quad |h_1|=|h_2|=...=|h_K|.\]
Therefore, it can be concluded that 
\[\Delta r_{O1}\leq \Delta r_N,\]
and the equality is achieved when 
\[|h_1|=|h_2|=...=|h_K|.\]
The proof of Theorem \ref{tm:2} is complete.
\end{IEEEproof}

\subsection{Solution of OMA-TYPE-II}
The superiority of NOMA compared to OMA-TYPE-II is shown in Theorem \ref{tm:3}.
\begin{theorem}\label{tm:3}
Given $P$ and $r^*$, if
\begin{equation}
\label{eqTM:P_O2^*}
P_{O2}^* \triangleq \min_{\sum_{i=1}^K \alpha_i=1} {N_0} \sum_{i=1}^{K }\frac{(e^{\frac{r^*}{\alpha_i}}-1)\alpha_i}{|h_{i}|^2} \leq P,
\end{equation}
then, the optimization problem in \eqref{eq:Opt_OMA_JO} is feasible. The optimal solution must satisfy
\begin{equation}
\label{eq:tm3}
R_{O2}\leq R_N,
\end{equation}
and the equality holds only when  $|h_1|=|h_2|=...=|h_K|$.
\end{theorem}
\begin{IEEEproof}
Again the total power is  split into two parts, i.e., the minimum power for supporting minimum rate transmission, and the excess power.

For user $i$, it is noted that the minimum power $P_i^*$ should satisfy
\[
\alpha_i\ln\Big(1+\frac{P_i^*|h_i|^2}{\alpha_i N_0}\Big)=r^*.
\] 
Hence, we can obtain
\[
P_i^*=N_0 \frac{(e^{\frac{r^*}{\alpha_i}}-1)\alpha_i}{|h_i|^2},
\]
and the total minimum power $P_{O2}^*$ can consequently be written as 
\begin{equation}
\label{eq:P_O2}
\begin{aligned}
P_{O2}^*&=\min_{\sum_{i=1}^K\alpha_i=1}\sum_{i=1}^KP_i^*\\
		&=\min_{\sum_{i=1}^K\alpha_i=1}N_0\sum_{i=1}^K \frac{(e^{\frac{r^*}{\alpha_i}}-1)\alpha_i}{|h_i|^2}.
\end{aligned}
\end{equation}

On the other hand, given user $i$, the rate increment with excess power $\Delta P_i$ can be calculated as
\begin{equation*}\label{eq:*1}
\begin{aligned}
\Delta r_i&=\alpha_i \ln\Big(1+\frac{(P_i^*+\Delta P_i)|h_i|^2}{\alpha_iN_0}\Big)-r^*\\
          &=\alpha_i \ln\Big(1+\frac{ \Delta P_i|h_i|^2}{\alpha_iN_0+ P_i^*|h_i|^2}\Big)\\
          &=\alpha_i \ln\Big(1+\frac{ \Delta P_i }{\alpha_iN_0 }\frac{|h_i|^2\alpha_i N_0}{\alpha_iN_0+P_i^*|h_i|^2}\Big)\\
          &=\alpha_i \ln\Big(1+\frac{ \Delta P_i }{\alpha_iN_0 }|h_i|^2e^{-\frac{r^*}{\alpha_i}}\Big).\\
\end{aligned}
\end{equation*}
By defining
\[
|\hat{h_i}|^2\triangleq  |h_i|^2 e^{-\frac{r^*}{\alpha_i}},
\]
the rate increment can be simply written as 
\[
\Delta r_i=\alpha_i\ln\Big(1+\frac{\Delta P_i|\hat{h_i}|^2}{\alpha_iN_0}\Big).
\]

Therefore, the optimization problem in \eqref{eq:Opt_OMA_JO} can be transformed to the problem as follows:
\begin{equation}\label{eq:Opt_OMA_eq} 
\begin{aligned}
R_{O2}= \max_{ {P_i},{\alpha_i}}\quad &Kr^*+\sum_{i=1}^K \Delta r_i\\
\mathrm{s.t.} \quad     &\sum_{i=1}^K P_i+ \sum_{i=1}^KP_i^*\leq P ,\\
                        &\Delta r_i=\alpha_i\ln \Big(1+\frac{ P_i|\hat{h_i}|^2}{\alpha_iN_0} \Big),\\
                        &\sum_{i=1}^K\alpha_i=1.
\end{aligned}
\end{equation}

Consequently, $R_{O2}$ can be written as 
\begin{equation}
\label{eq:OMA_sol}
R_{O2}= Kr^*+\Delta r_{O2},
\end{equation}
where
\begin{equation}\label{eq:Delta_r_O2} 
\begin{aligned}
\Delta r_{O2}=\max_{P_i,\alpha_i}\quad &\sum_{i=1}^K\Delta r_i\\
\mathrm{s.t.}\quad &\sum_{i=1}^KP_i+N_0\sum_{i=1}^K\frac{(e^{\frac{r^*}{\alpha_i}}-1)\alpha_i}{|h_i|^2}\leq P ,\\
&\Delta r_i=\alpha_i\ln \Big(1+\frac{ P_i|{h_i}|^2}{\alpha_ie^\frac{r^*}{\alpha_i}N_0}\Big),\\
                   &\sum_{i=1}^K\alpha_i=1.
\end{aligned}
\end{equation}
It is worth noting that the optimization problem in \eqref{eq:Delta_r_O2} is non-convex, and finding the a closed-form expression for its optimum solution  or a good  upper bound is very difficult. For example, if one uses Jensen's inequality on the objective function as we have done before, it will lead to meaningless results, which will be explained in the following.
 
By using Jensen's inequality, we have
\begin{equation}\label{eq:r_O2_Jesen}
\begin{aligned}
&\sum_{i=1}^K\alpha_i\ln \Big(1+\frac{ P_i|\hat{h_i}|^2}{\alpha_iN_0}\Big)\\
&\leq \ln\Big( \sum_{i=1}^K \alpha_i\big(1+\frac{P_i|\hat{h_i}|^2}{\alpha_iN_0}\big)\Big)\\
&=\ln\Big(   1+\sum_{i=1}^K\frac{P_i|\hat{h_i}|^2}{ N_0} \Big).
\end{aligned}
\end{equation}
By combining \eqref{eq:Delta_r_O2} and \eqref{eq:r_O2_Jesen}, we can obtain
\begin{equation}\label{eq:rO2_upper} 
\begin{aligned}
\Delta r_{O2}\leq \max_\mathcal{P}\quad & \ln \Big(1+\sum_{i=1}^K \frac{P_i|\hat{h_i}|^2}{N_0}\Big)\\
\mathrm{s.t.}\quad &\sum_{i=1}^KP_i\leq P-P_{O2}^* .
\end{aligned}
\end{equation}
The optimal solution of the optimization problem in \eqref{eq:tm3_rO_upper} is to allocate all the power to user 1, i.e., $P_1=P-P_{O2}^*$.
Therefore, we can have
\begin{equation}\label{eq:TM3_a1} 
\begin{aligned}
\Delta r_{O2}&\leq \ln \Big(1+  \frac{(P-P_{O2}^*)|\hat{h_1}|^2}{N_0}\Big)\\
			 &= \ln \Big(1+  \frac{(P-P_{O2}^*)| {h_1}|^2}{e^{\frac{r^*}{\alpha_i}}N_0}\Big)\\
			 &\leq \ln \Big(1+  \frac{(P-P_{O2}^*)| {h_1}|^2}{e^{r^*}N_0}\Big).
\end{aligned}
\end{equation}
Obviously, this upper bound is too loose to be meaningful.

To derive a tighter upper bound for the optimization problem in \eqref{eq:Delta_r_O2}, we introduce the following Lemma.
\begin{lemma}[Upper bound for $\Delta r_{O2}$]\label{lm:2}
The optimal solution of  \eqref{eq:Delta_r_O2} can be upper-bounded by
\[
\Delta r_{O2} \leq \ln \Big(1+  \frac{(P-P_{O2}^*)| {h_1}|^2}{e^{Kr^*}N_0}\Big).
\]
\end{lemma} 
\begin{IEEEproof}
We can rewrite $P_i$ as follows:
\begin{equation}\label{eq:lm:2_1}
P_i=N_0\frac{\alpha_ie^{\frac{r^*}{\alpha_i}}}{|h_i|^2}(e^{\frac{\Delta r_i}{\alpha_i}}-1).
\end{equation}
Then, we can obtain
\begin{equation}\label{eq:lm:2_2}
\begin{aligned}
N_0\sum_{i=1}^N\frac{1}{|h_i|^2}(e^{\frac{r^*+\Delta r_i}{\alpha_i}}-1)\alpha_i\leq P.
\end{aligned}
\end{equation}
By recalling the definition of $P_{O2}^*$ in \eqref{eq:P_O2}, we can have
\begin{equation}\label{eq:lm:2_3}
\begin{aligned}
P-P_{O2}^*&\geq  
N_0\sum_{i=1}^N\frac{1}{|h_i|^2}(e^{\frac{r^*+\Delta r_i}{\alpha_i}}-1)\alpha_i\\
&-\min_{\sum_{i=1}^K\alpha_i=1} N_0\sum_{i=1}^N\frac{1}{|h_i|^2}(e^{\frac{r^* }{\alpha_i}}-1)\alpha_i.
\end{aligned}
\end{equation}
Denote
\[
\frac{1}{|h_i|^2}=\frac{1}{|h_1|^2}+\Delta g_i, \quad i=2,3,...,K.
\]
The right hand of \eqref{eq:lm:2_3} can be further lower-bounded by
\begin{equation}\label{eq:lm:2_4}
\begin{aligned}
&N_0\sum_{i=1}^K\frac{1}{|h_i|^2}(e^{\frac{r^*+\Delta r_i}{\alpha_i}}-1)\alpha_i
-\min_{\sum_{i=1}^K\alpha_i=1} N_0\sum_{i=1}^K\frac{1}{|h_i|^2}(e^{\frac{r^* }{\alpha_i}}-1)\alpha_i\\
&=\frac{N_0}{|h_1|^2}\sum_{i=1}^K(e^{\frac{r^*+\Delta r_i}{\alpha_i}}-1)\alpha_i
+N_0\sum_{i=2}^Kg_i(e^{\frac{r^*+\Delta r_i}{\alpha_i}}-1)\alpha_i\\
&-\min_{\sum_{i=1}^K\alpha_i=1}\Big(\frac{N_0}{|h_1|^2}\sum_{i=1}^K(e^{\frac{r^* }{\alpha_i}}-1)\alpha_i
+N_0\sum_{i=2}^Kg_i(e^{\frac{r^* }{\alpha_i}}-1)\alpha_i \Big)\\
&\geq \frac{N_0}{|h_1|^2}\sum_{i=1}^K(e^{\frac{r^*+\Delta r_i}{\alpha_i}}-1)\alpha_i
-\frac{N_0}{|h_1|^2}\min_{\sum_{i=1}^K\alpha_i=1} \sum_{i=1}^K(e^{\frac{r^* }{\alpha_i}}-1)\alpha_i\\
&\geq \frac{N_0}{|h_1|^2}\big(e^{Kr^*+\sum_{i=1}^K\Delta r_i}-1\big)-\frac{N_0}{|h_1|^2}\big(e^{Kr^*}-1\big)\\
&=\frac{N_0}{|h_1|^2}e^{Kr^*}\big(e^{ \sum_{i=1}^K\Delta r_i}-1 \big),
\end{aligned}
\end{equation}
where the last inequality holds because of Jensen's inequality on the convex function $f(x)=e^{x}-1$.
Consequently, by combining \eqref{eq:lm:2_3} and \eqref{eq:lm:2_4}, we finally obtain that
\begin{equation}
\label{eq:lm:2_5}
\frac{N_0}{|h_1|^2}e^{Kr^*}\big(e^{ \sum_{i=1}^K\Delta r_i}-1 \big)\leq P-P_{O2}^*.
\end{equation}
Therefore, we have
\begin{equation}
\label{eq:lm:2_6}
\Delta r_{O2}=\sum_{i=1}^K\Delta r_i\leq \ln\Big(1+\frac{(P-P_{O2}^*)|h_1|^2}{e^{Kr^*}N_0} \Big),
\end{equation}
and  Lemma \ref{lm:2} is proved.
\end{IEEEproof}

To prove Theorem \ref{tm:3}, we also need another lemma given next to characterize the lower bound of $P_{O2}^*$.
\begin{lemma}[Lower bound for $P_{O2}^*$]\label{lm:3}
The lower bound of $P_{O2}^*$ is $P_N^*$, i.e.,
\[
P_{O2}^*\geq P_N^*.
\]
\end{lemma} 
\begin{IEEEproof}
By recalling the definition of $P_N^*$ and  $P_{O2}^*$ in Theorems \ref{tm:1} and \ref{tm:3}, it is noticed that $P_{O2}^*$ can also be written as follows:
\begin{equation*}
\label{eq: }
\begin{aligned}
P_{O2}^*
		&=\min_{\sum_{i=1}^K\alpha_i=1}N_0\sum_{i=1}^K \frac{(e^{\frac{r^*}{\alpha_i}}-1)\alpha_i}{|h_i|^2}\\
		&=\min_{\sum_{i=1}^K\alpha_i\leq 1}N_0\sum_{i=1}^K \frac{(e^{\frac{r^*}{\alpha_i}}-1)\alpha_i}{|h_i|^2},
\end{aligned}
\end{equation*}
the second inequality holds since the objective function is minimized only when $\sum_{i=1}^K\alpha_i=1$ holds (complementary slackness). 
We only need to prove that the following inequality
\begin{equation}
\label{eq:lm:3_1}
{N_0} \sum_{i=1}^{K }\frac{(e^{\frac{r^*}{\alpha_i}}-1)\alpha_i}{|h_{i}|^2}\geq (e^{r^*}-1)N_0\sum_{i=0}^{K-1}\frac{e^{ir^*}}{|h_{K-i}|^2}
\end{equation}
holds for all $\alpha_i$ satisfying $\sum_{i=1}^K\alpha_i\leq 1$.
This inequality can also be simplified as follows:
\begin{equation}
\label{eq:lm:3_2}
\sum_{i=1}^{K }\frac{1}{{|h_{i}|^2}}\frac{(e^{\frac{r^*}{\alpha_i}}-1)\alpha_i}{(e^{r^*}-1)}\geq  \sum_{i=1}^{K }\frac{1}{{|h_{i}|^2}}e^{(K-i)r^*}.
\end{equation}
To prove \eqref{eq:lm:3_2}, we first introduce the following lemma.
\begin{lemma}[]\label{lm:4}
Given 
$
0<c_1\leq c_2\leq ... \leq c_K,
$
if 
$\sum_{i=j}^{K }a_i\geq \sum_{i=j}^{K }b_i$
holds for $j=1,2,...,K$,
then, we can have
\[
\sum_{i=1}^Kc_ia_i\geq \sum_{i=1}^{K}c_ib_i.
\]
\end{lemma} 
\begin{IEEEproof}
We first define a non-negative sequence $d_j,\quad j=1,2,3,...,K$ as follows:
\[d_1=c_1, \quad d_j=c_j-\sum_{i=1}^{j-1}d_i, \quad j=2,3,...,K.\]
Since $\sum_{i=j}^{K }a_i\geq \sum_{i=j}^{K }b_i$
holds for $j=1,2,...,K$, we have the following $K$ inequalities.
\begin{equation}\label{eq:lm4_1}
\left\{
\begin{aligned}
a_1+a_2+...+a_K&\geq b_1+b_2+...+b_K\\
a_2+...+a_K&\geq b_2+...+b_{K}\\
&...\\
a_K&\geq b_K.\\
\end{aligned}
\right. 
\end{equation}
By multiplying $d_i, i=1,2,...,K$ with the $K$ inequalities in \eqref{eq:lm4_1} respectively and adding them together, we can have
\[
\sum_{i=1}^Kc_ia_i\geq \sum_{i=1}^{K}c_ib_i,
\]
and   Lemma \ref{lm:4} is proved.
\end{IEEEproof}

By defining 
\begin{equation}\label{eq:lm:3_3}
\left\{
\begin{aligned}
&a_i=\frac{(e^{\frac{r^*}{\alpha_i}}-1)\alpha_i}{e^{r^*}-1},\\
&b_i=e^{(K-i)r^*},
\end{aligned}
\right. 
\end{equation}
we can easily check that
\[
\sum_{i=j}^{K }a_i\geq \sum_{i=j}^{K }b_i
\]
holds for $\sum_{i=1}^K\alpha_i\leq 1$.
Therefore, by taking advantage of Lemma \ref{lm:4}, \eqref{eq:lm:3_2} can be obtained, and Lemma \ref{lm:3} is proved. 
\end{IEEEproof}

By combining Lemmas \ref{lm:2} and \ref{lm:3}, we can finally conclude that
\begin{equation}
\label{eq:**}
\Delta r_{O2}\leq \ln\Big(1+\frac{(P-P_{N}^*)|h_1|^2}{e^{Kr^*}N_0} \Big),
\end{equation}
and the proof is completed.
\end{IEEEproof}

\subsection{Major Results}
The major analytical results of this paper can be summarized in the following.
\begin{enumerate}[(1)]
\item To support reliable data transmission with minimum rate constraint, the required minimum powers of NOMA, OMA-TYPE-I, and OMA-TYPE-II can be written as follows.
\begin{equation}\label{eq:Con1}
\left\{
\begin{aligned}
&P_N^* = (e^{r^*}-1)N_0\sum_{i=0}^{K-1}\frac{e^{ir^*}}{|h_{K-i}|^2},\\
&P_{O1}^* \triangleq (e^{Kr^*}-1)\frac{N_0}{K}\sum_{i=1}^{K }\frac{1}{|h_{i}|^2},\\
&P_{O2}^* \triangleq \min_{\sum_{i=1}^K \alpha_i=1} {N_0} \sum_{i=1}^{K }\frac{(e^{\frac{r^*}{\alpha_i}}-1)\alpha_i}{|h_{i}|^2}.\\
\end{aligned}
\right. 
\end{equation}
\item The relationship of the required minimum powers of NOMA, OMA-TYPE-I, and OMA-TYPE-II are
\begin{equation}
\label{eq:Con2}
P_N^*\leq P_{O2}^*\leq P_{O1}^*.
\end{equation}
\item The closed-form expression for the optimum sum rate of NOMA systems can be written as 
\begin{equation}
\label{eq:Con3}
R_N=Kr^*+\ln\Big(1+\frac{(P-P_N^* )|{h_1}|^2}{N_0e^{Kr^*}}\Big).
\end{equation}
\item The optimum sum rates of NOMA, OMA-TYPE-I, and OMA-TYPE-II have the following relationship
\begin{equation}
\label{eq:Con4}
R_N\geq R_{O2}\geq R_{O1}.
\end{equation}
\end{enumerate}

\section{Simulation Results}
In this section, computer simulations are conducted  to validate the correctness of the analytical results. 
The signal-to-noise ratio (SNR) is defined as $\mathrm{SNR}=10\log{\frac{P}{N_0}}$.
Simulation results in this section are given for both deterministic channels and Rayleigh fading channels. Particularly, numerical examples based on deterministic channels are given first to validate our analytical conclusions and then the results based on Rayleigh fading channels are given  to offer more insights about the differences among NOMA, OMA-TYPE-I and OMA-TYPE-II systems.
\subsection{Deterministic Channels}
Since the mathematical analysis in this paper is based on deterministic channels, we first validate our analytical results by the following numerical investigations with fixed channel realizations.

\begin{figure}[h]
\centering
\includegraphics[width=\figa ]{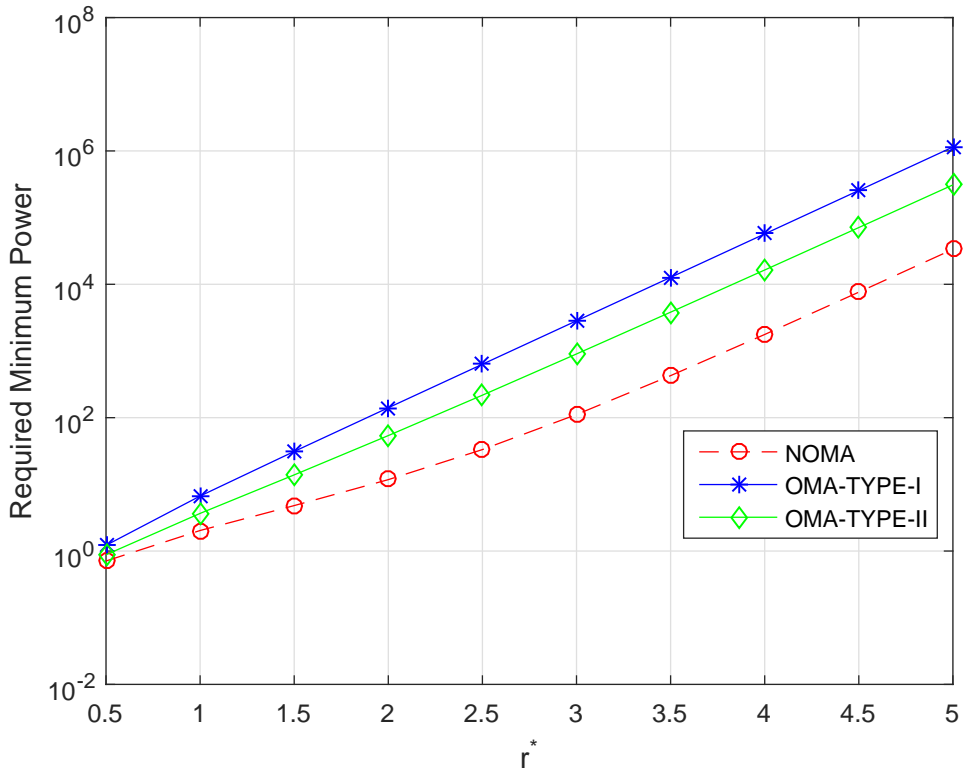} 
\caption{Required minimum power  versus $r^*$ for $K=3$ users, with $(h_1,h_2,h_3)=(10,5,1).$}
\label{fig:1}
\end{figure}
\begin{figure}[h]
\centering
\includegraphics[width=\figa ]{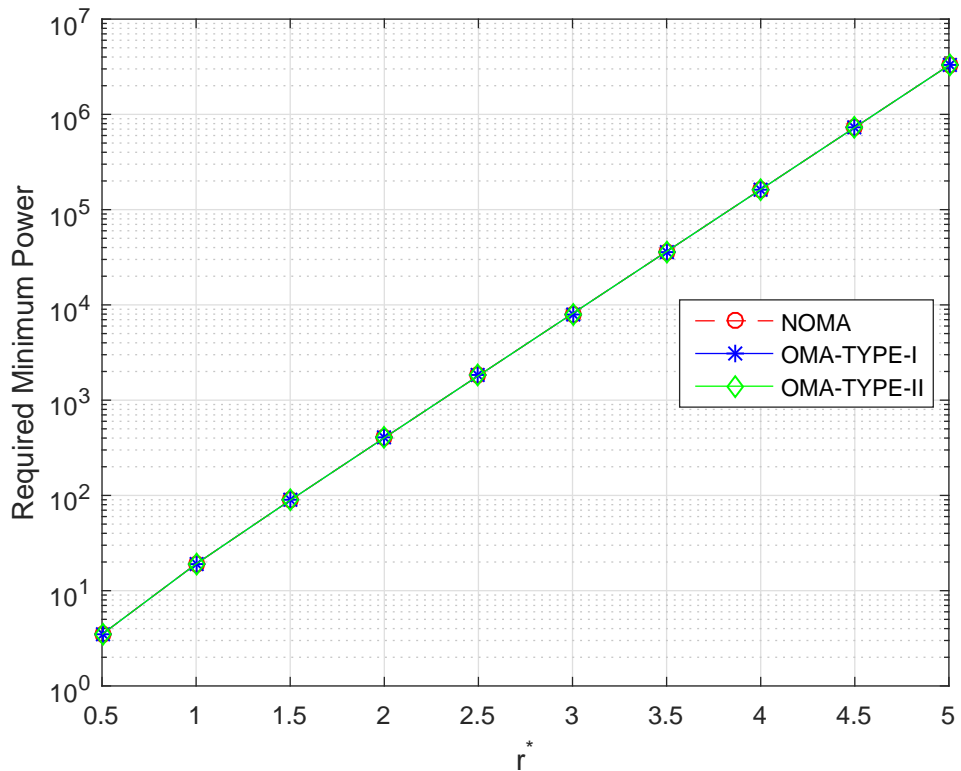} 
\caption{Required minimum power  versus $r^*$ for $K=3$ users, with $(h_1,h_2,h_3)=(1,1,1).$}
\label{fig:2}
\end{figure}
Figs. \ref{fig:1} and \ref{fig:2} show  the required minimum power versus the target  minimum rate $r^*$ for different transmission schemes given  specified channel realizations.  
The required minimum power for NOMA, OMA-TYPE-I and OMA-TYPE-II is obtained by the analytical results in \eqref{eq:Con1}.
In Fig. \ref{fig:1}, the channel coefficients are fixed to be $(h_1,h_2,h_3)=(10,5,1)$,
and in Fig. \ref{fig:2}, the channel coefficients are fixed to be identical, i.e., $(h_1,h_2,h_3)=(1,1,1)$.
By observing these two figures, we have the following comments.
\begin{enumerate}
\item All the required minimum power of the three systems, i.e., $P_N^*,P_{O1}^*,P_{O2}^*$,  increases exponentially as the target minimum rate, i.e., $r^*$, increases.
\item When the channel coefficients are not the same, the required minimum power of OMA-TYPE-II is smaller than that of OMA-TYPE-I, while the required minimum power of NOMA is smaller than that of OMA-TYPE-II. Note that these observations are consistent with our analytical results in \eqref{eq:Con2}.
\item When the channel coefficients are identical, all the three kinds of required minimum power become  the same.
\end{enumerate}

\begin{figure}[h]
\centering
\includegraphics[width=\figa ]{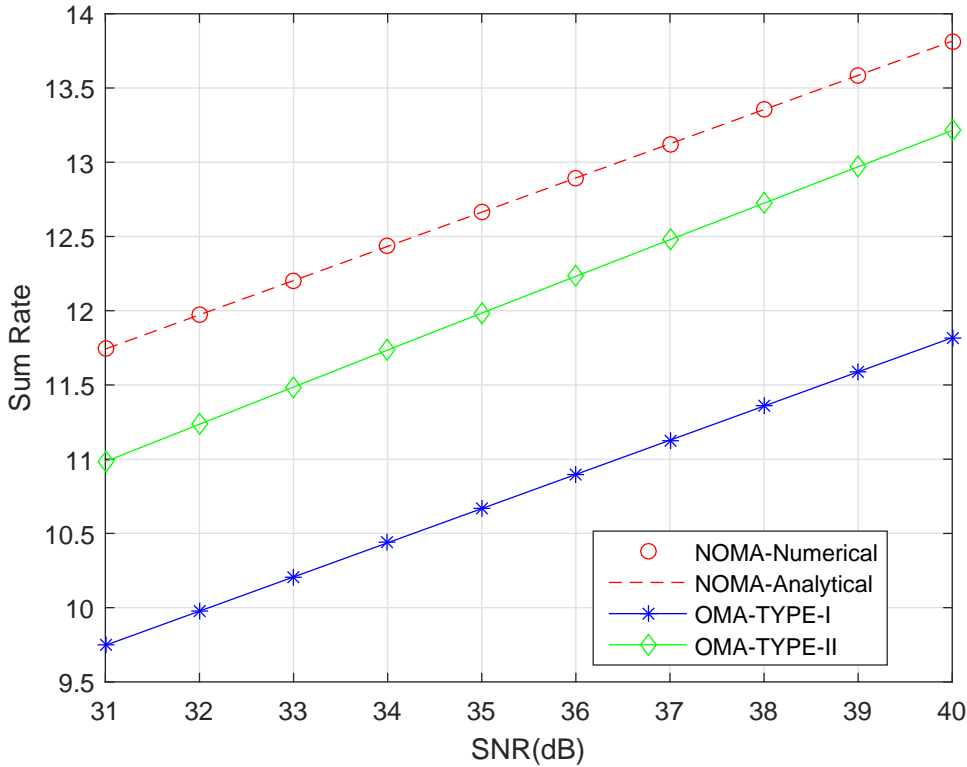}
\caption{Sum rates  versus SNR, with $K=3$, and $(h_1,h_2,h_3,r^*)=(10,5,1,1)$}
\label{fig:3}
\end{figure}
\begin{figure}[h]
\centering
\includegraphics[width=\figa ]{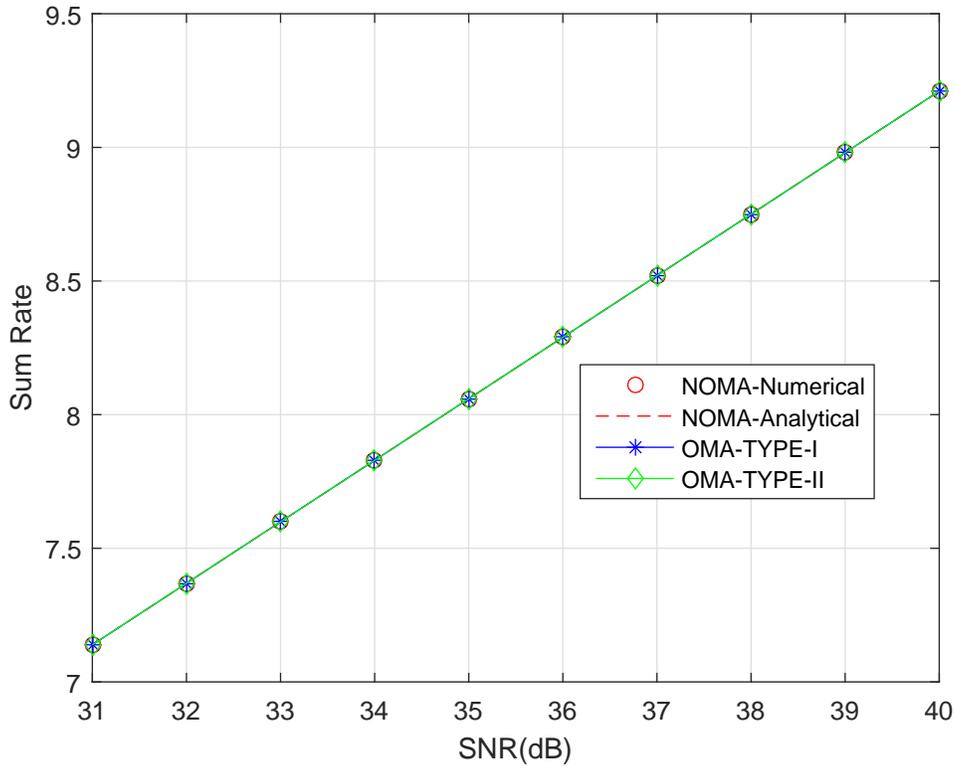}
\caption{Sum rates  versus SNR, with $K=3$, and $(h_1,h_2,h_3,r^*) =(1,1,1,1)$}
\label{fig:4}
\end{figure}
Figs. \ref{fig:3} and \ref{fig:4} show  the sum rates versus SNR for different transmission schemes given  specified channel realizations. 
The optimum sum rates for NOMA-Numerical, OMA-TYPE-I, and OMA-TYPE-II are obtained by solving the optimization problems in \eqref{eq:Opt_NOMA}, \eqref{eq:Opt_OMA} and \eqref{eq:Opt_OMA_JO}, respectively. The optimum sum rates for NOMA-Analytical are attained by the analytical closed-form expression in \eqref{eq:Con3}.
In Fig. \ref{fig:3}, the channel coefficients are fixed to be $(h_1,h_2,h_3)=(10,5,1)$,
and in Fig. \ref{fig:4}, the channel coefficients are fixed to be identical, i.e., $(h_1,h_2,h_3)=(1,1,1)$. For both channel setups, we set $r^*=1$. By observing these figures, we have the following comments.
\begin{enumerate}
\item The numerical and analytical results for NOMA match perfectly.
\item When the channel coefficients are not the same, the sum rates of OMA-TYPE-II are always larger than those of OMA-TYPE-I, while the sum rates of NOMA are always larger than those of OMA-TYPE-II. Note that these observations are also consistent with our analytical results in \eqref{eq:Con4}.
\item When the channel coefficients are identical, all the three kinds of sum rates become the same.
\end{enumerate}

\subsection{Rayleigh Fading Channels}
With randomly generated wireless channels, e.g., Rayleigh fading channels, herein, we introduce two performance evaluation metrics, e.g., outage probability and ergodic sum  rate, to evaluate and compare the performance of NOMA, OMA-TYPE-I and OMA-TYPE-II.

Recall that a system is in outage if there exists one user who cannot receive its own messages with the given target minimum rate $r^*$ for all the possible resource allocation, i.e., the corresponding optimization problem is infeasible. Mathematically, the outage probabilities of NOMA, OMA-TYPE-I and OMA-TYPE-II can be written as
\[
\mathrm{Pr}\{P_N^*>P\},\quad \mathrm{Pr}\{P_{O1}^*>P\},\quad \mathrm{Pr}\{P_{O2}^*>P\},
\]
respectively, and this criterion  will be used in Figs. 5 and 6.
 
{
In Figs. 7 and 8, we will use the ergodic sum rate as the criterion to evaluate the performance of NOMA, OMA-TYPE-I and OMA-TYPE-II. 
These ergodic sum rates can be defined as  in the following.
Without loss of generality, take NOMA system  as an example.
Denote $R_N(\mathbf{h})$ by the instantaneous optimum sum rate achieved by NOMA and $P_N^*(\mathbf{h})$ by the required minimum power of NOMA given  a specific channel realization  $\mathbf{h}=[h_1,h_2,...,h_K]^T$. 
Note that the instantaneous optimum sum rate reduces to zero if the optimization problem in \eqref{eq:Opt_NOMA} is infeasible, i.e., the system is in outage.
Therefore, the instantaneous optimum sum rate achieved by NOMA can be mathematically expressed as follows:
\begin{equation*}
R_N(\mathbf{h})=
\left\{
\begin{aligned}
&R_N \quad &\text{if} \quad P_N^*(\mathbf{h})\leq P, \\
&0\quad &\text{Otherwise,}
\end{aligned}
\right. 
\end{equation*}
where $R_N$ and $P_N^*(\mathbf{h})$ are defined in \eqref{eq:Con1} and \eqref{eq:Con3}, respectively.
With such a definition of the instantaneous optimum sum rate, the ergodic sum rate of NOMA is defined as the expectation of $R_N(\mathbf{h})$ with respect to independent and identically distributed (i.i.d.) Rayleigh fading $h_i$'s .
Note that the ergodic sum rates of OMA-TYPE-I and OMA-TYPE-II can be defined similarly.
}
%

\begin{figure}[h]
\centering
\includegraphics[width=\figa ]{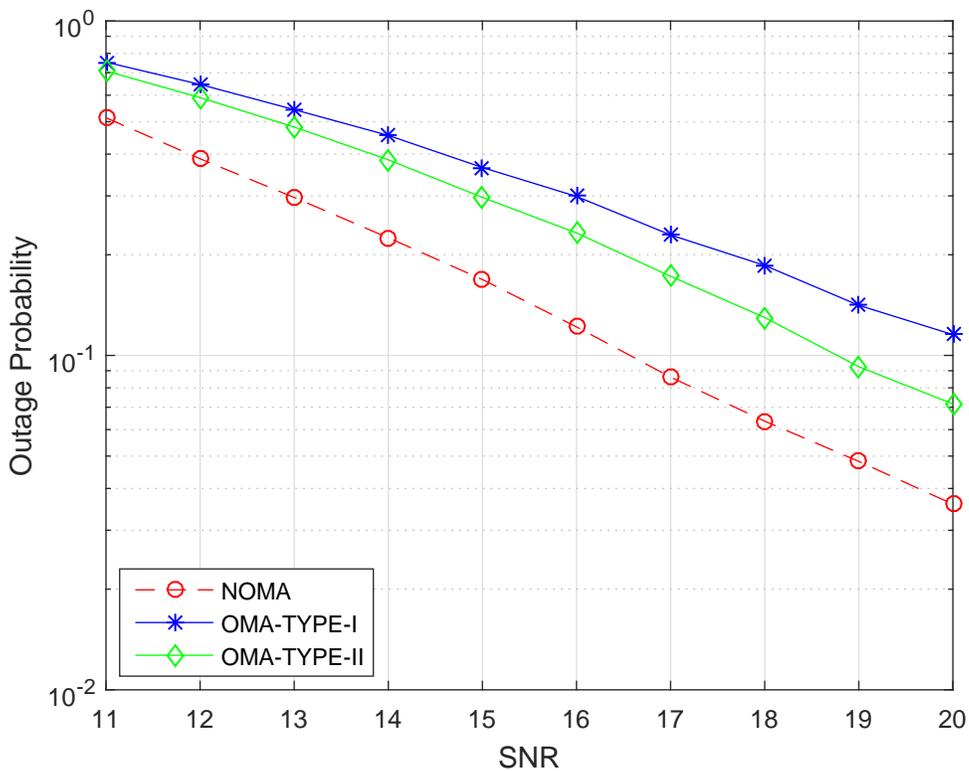} 
\caption{Outage Probability  versus SNR, with $K=3,r^*=1$}
\label{fig:5}
\end{figure}
\begin{figure}[h]
\centering
\includegraphics[width=\figa ]{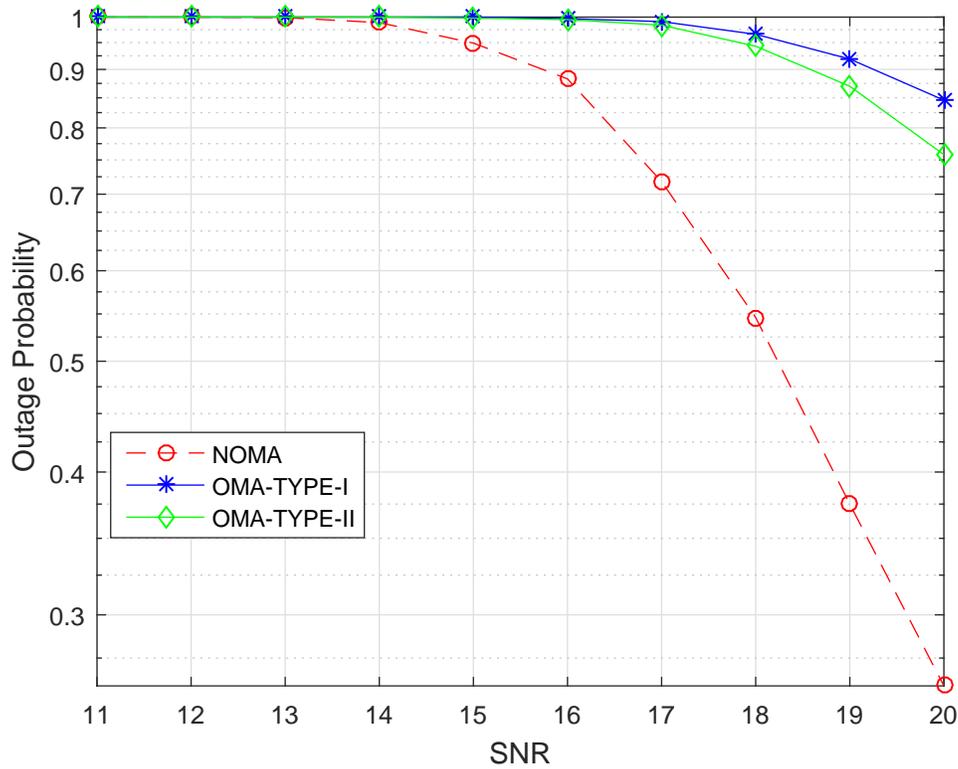} 
\caption{Outage Probability  versus SNR, with $K=5,r^*=1$}
\label{fig:6}
\end{figure}
In Figs. \ref{fig:5} and \ref{fig:6}, given a fixed target minimum rate $r^*=1$, the outage performance versus the SNR for different transmission schemes under Rayleigh fading channels are plotted with $K=3$ and $K=5$, respectively. 
Since our analytical results show that $P_N^*\leq P_{O2}^*\leq P_{O1}^*$, we can infer that
\[
\mathrm{Pr}\{P_N^*>P\}\leq \mathrm{Pr}\{P_{O2}^*>P\}\leq \mathrm{Pr}\{P_{O1}^*>P\}.
\]
This conclusion is confirmed by both Figs. \ref{fig:5} and \ref{fig:6}. Particularly, in Fig. \ref{fig:5}, OMA-TYPE-II yields about a gain of 1.5dB  over OMA-TYPE-I, and NOMA has about a gain of 2.5dB  over OMA-TYPE-II at $\mathrm{Pr}=10^{-1}$. Moreover, by comparing Fig. \ref{fig:5} with Fig. \ref{fig:6}, it is also observed that the outage probability gain by NOMA becomes larger when the number of users increases.

\begin{figure}[h]
\centering
\includegraphics[width=\figa ]{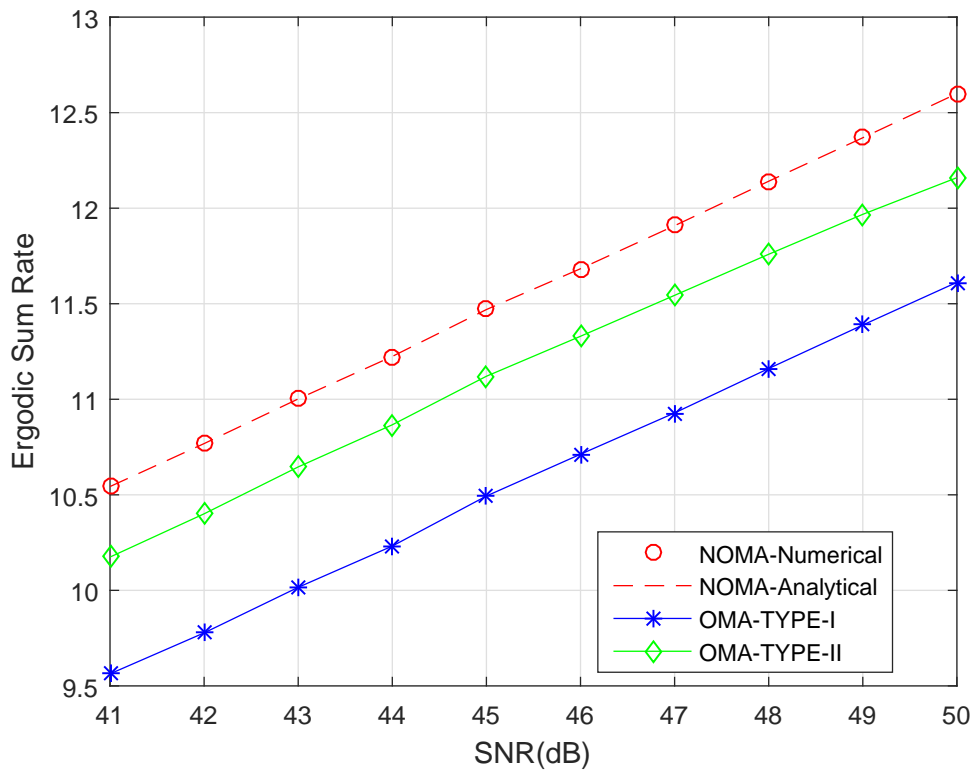} 
\caption{Ergodic Sum Rates  versus SNR, with $K=3, r^*=1$}
\label{fig:7}
\end{figure}
\begin{figure}[h]
\centering
\includegraphics[width=\figa ]{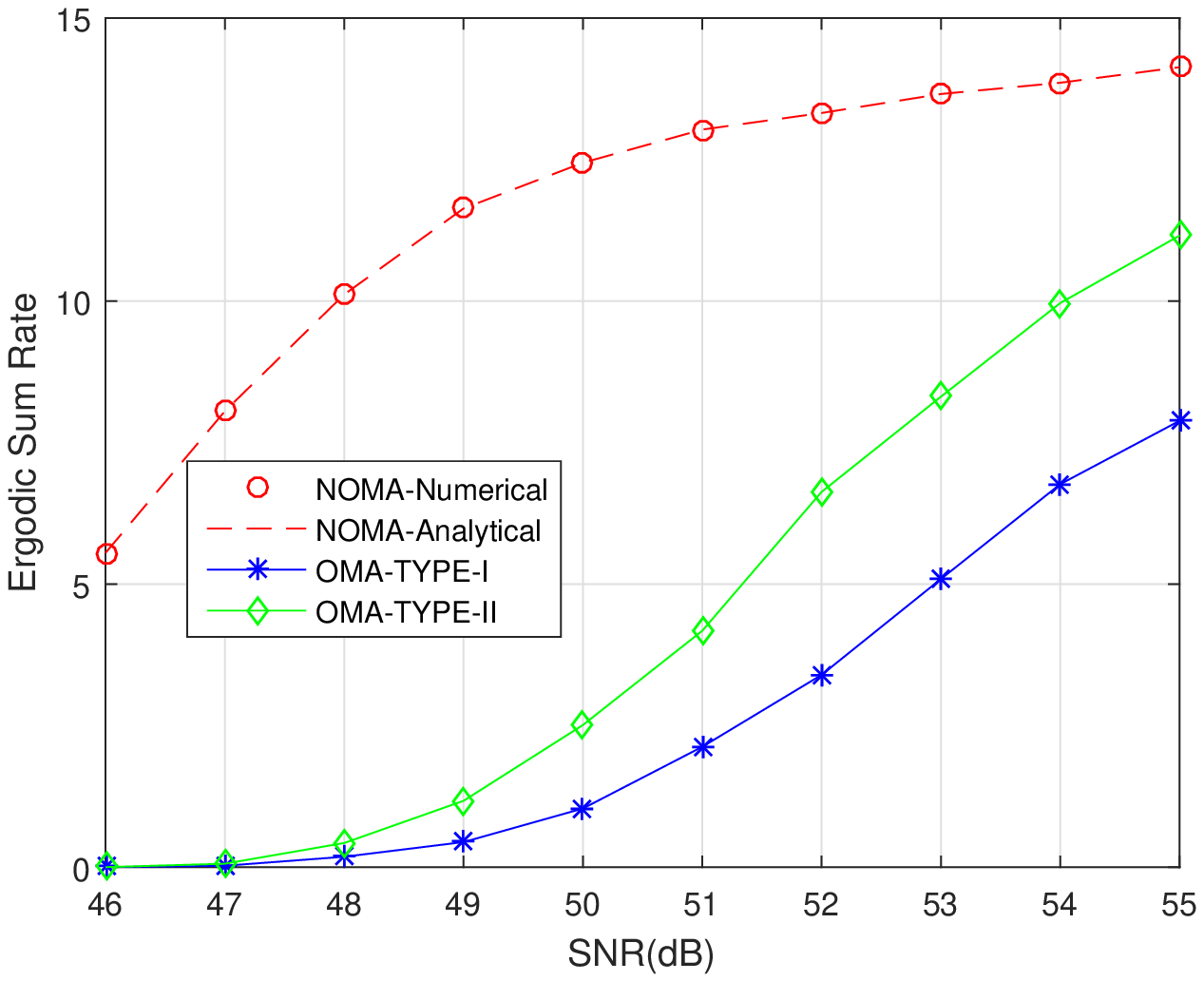} 
\caption{Ergodic Sum Rates  versus SNR, with $K=6, r^*=2$}
\label{fig:8}
\end{figure}

In Figs. \ref{fig:7} and \ref{fig:8}, the ergodic sum rate performance versus SNR for different transmission schemes under Rayleigh fading channels are plotted with $K=3,r^*=1$ and $K=6,r^*=2$, respectively. 
By observing these figures, we have the following comments.
\begin{enumerate}
\item The ergodic sum rate of NOMA is always larger than that of OMA-TYPE-II, and the ergodic sum rate of OMA-TYPE-II is always larger than that of OMA-TYPE-I.
\item When the transmission power is large enough with respect to the target minimum rate $r^*$,  i.e., the outage probabilities for all the three systems tend  to  zero, the ergodic sum rate increases linearly with SNR. For example, in Fig. \ref{fig:7}, NOMA has about a gain of 0.3 nats per channel use (NPCU) over OMA-TYPE-II, and OMA-TYPE-II has about a gain of  0.7 NPCU over OMA-TYPE-I, for all the SNRs.
\item When the transmission power is not large enough with respect to the target minimum rate $r^*$, i.e., a system may be in outage, both OMA-TYPE-I and OMA-TYPE-II may suffer a significant performance loss compared  to NOMA in the low SNR regime. For example, in Fig. \ref{fig:8}, when $\mathrm{SNR}=46 \mathrm{dB}$, the ergodic sum rates of OMA-TYPE-I and OMA-TYPE-II decease to nearly zero, while the ergodic sum rate of NOMA can be still maintained over 5 NPCU. 
\end{enumerate}

\section{Conclusion}
In this paper, we have mathematically compared the optimum sum rate performance for NOMA and OMA systems, with consideration of user fairness. 
Firstly, the closed-form optimum sum rate and the corresponding power allocation policy for NOMA systems have been derived, by using the power splitting method. 
Secondly, the fact that NOMA can always achieve better sum rate performance than that of  traditional OMA-TYPE-I with optimum power allocation but equal user time/frequency allocation  has been revealed, by a rigorous mathematical proof.
Thirdly, we have  proved that NOMA can also outperform OMA-TYPE-II with power and time/frequency allocation jointly optimized in terms of sum rate performance. 
Moreover, the major analytical results have been extracted from those mathematical proofs.
Finally, computer simulations have been conducted to validate the correctness of these analytical results and show the advantages of NOMA over OMA in practical Rayleigh fading channels.

\bibliographystyle{IEEEtran}
\bibliography{IEEEabrv,Letter}
\end{document}